\documentclass[12pt]{article}%
\usepackage{amsfonts}
\usepackage{amsmath}
\usepackage{amssymb}
\usepackage{graphicx}%
\begin{document}
\begin{titlepage}
\vspace{1.5cm}
\begin{center}
\
\\
{\bf\large  Generalized screened potential model }
\\
\date{ }
\vskip 0.70cm
P. Gonz\'{a}lez
\vskip 0.30cm
{ \it Departamento de F\'{\i}sica Te\'orica -IFIC\\
Universidad de Valencia-CSIC \\
E-46100 Burjassot (Valencia), Spain.} \\ ({\small E-mail:
pedro.gonzalez@uv.es})
\end{center}
\vskip 1cm \centerline{\bf Abstract}
A new non relativistic quark model to calculate the spectrum of heavy quark
mesons is developed. The model is based on an interquark potential interaction
that implicitly incorporates screening effects from meson-meson
configurations. An analysis of the bottomonium spectrum shows the appearance
of extra states as compared to conventional non screened potential models.
\end{titlepage}
\section{Introduction\label{SI}}
Non Relativistic Quark Models (NRQM) of hadron structure are based on the
consideration of effective quark degrees of freedom bound by an interquak
interaction potential. For heavy quark mesons $\left(  Q\overline{Q}\right)
$, in particular bottomonium $\left(  Q=b\right)  $ to which we shall restrict
our attention, the form of this potential can be inferred from lattice QCD.
More precisely, one can calculate in the lattice $E\left(  r\right)  $, the
energy of two static color sources, $Q$ and $\overline{Q}$, in terms of the
$Q-\overline{Q}$ distance. By identifying $E\left(  r\right)  $ with the sum
of the masses of the Quark $\left(  m_{Q}\right)  $ and the antiQuark $\left(
m_{\overline{Q}}\right)  $ plus the $Q\overline{Q}$ potential $V(r)$ one gets
$V(r)=E(r)-m_{Q}-m_{\overline{Q}}.$
\bigskip
$E\left(  r\right)  $ is calculated in the lattice from a correlation
function. In the so called quenched approximation (only the bare valence
$Q_{0}\overline{Q}_{0}$ configuration) a Cornell type of potential,
$V_{0}(r)=\sigma r-\frac{\chi}{r}+E_{0}$ where $\sigma,$ $\chi$ and $E_{0}$
are constants, comes out (see for example \cite{Bal01}). This simple potential
form provides a reasonable overall description of the masses of the low lying
heavy quarkonia states \cite{Eic80} although some refinements are needed to
reach a precise fit (see for example \cite{GI85}).
\bigskip
The form of $E\left(  r\right)  $ is altered when sea quarks are also taken
into account (unquenched lattice calculation). Actually the presence of
$q\overline{q}$ pairs where $q$ stands for a light quark $\left(
q=u,d,s\right)  $ gives rise to a screening of the color charges of the bare
valence quarks $Q_{0}$ and $\overline{Q}_{0}.$ A potential parametrization of
this effect was first proposed in the late eighties from a lattice calculation
in the two color case including dynamic Kogut-Susskind fermions and with a
lattice spacing fixed from the $\rho$ mass \cite{Bor89}. The resulting
Quark-antiQuark screened potential form was used, with parameters fixed
phenomenologically, to analyze heavy quark mesons as $Q\overline{Q}$ bound
states, $Q$ standing for an effective quark \cite{Din95,Gon03}. However, more
recent lattice data \cite{Bal05} that take into consideration interacting bare
valence $Q_{0}\overline{Q}_{0}$ and meson $\left(  Q_{0}\overline{q}\right)  $
- meson $\left(  \overline{Q}_{0}q\right)  $ configurations (see next
Section), suggest that the form of the potential should be different for
energies below and above a meson - meson threshold.
\bigskip
In this article we try to go a step further in the analysis of the heavy quark
meson spectrum within a non relativistic quark model framework by proposing
the form that a generalized interquark potential incorporating screening may
have below and above a meson-meson threshold. This proposal is based on the
assumption that screening effects are mainly due to the formation of meson
$\left(  Q_{0}\overline{q}\right)  $ - meson $\left(  \overline{Q}%
_{0}q\right)  $ structures since mesons are color singlets. Then we use the
lattice results for the static interquark energy $E\left(  r\right)  ,$ when
the bare valence quark and meson-meson configurations are considered
altogether, to build the screened potential.
\bigskip
The contents of the article are organized as follows. In Section \ref{SII} a
brief review of the lattice results for $E\left(  r\right)  $ is presented.
From them a generalized screened potential is defined. The resulting model is
applied, in Section \ref{SIII}, to calculate the bottomonium spectrum and to
analyze the spectral effect of screening by comparing the masses obtained with
the ones calculated from a non screened Cornell potential. Finally in Section
\ref{SIV} our main results and conclusions will be summarized.
\section{Generalized Screened Potential Model (GSPM)\label{SII}}
\begin{figure}[ptb]
\begin{center}
\includegraphics[height=2.5131in,width=2.8764in]{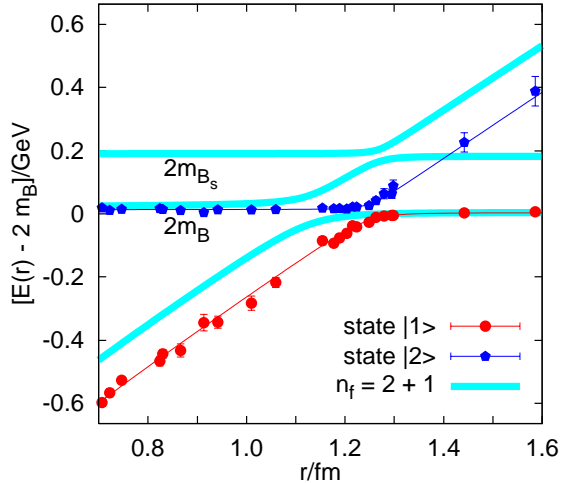}
\caption{Calculated $b\overline{b}$ energy from lattice QCD when a
$B\overline{B}$ configuration with mass $2m_{B}$ is implemented: circles and
pentagons over the thin lines. Educated guess for the case of two meson-meson
configurations, $B\overline{B}$ with mass $2m_{B}$ and $B_{s}\overline{B}_{s}$
with mass $2m_{B_{s}}$: thick lines. From reference \cite{Bal05}.}%
\label{Fig22prd05}%
\end{center}
\end{figure}

\begin{figure}
[ptb]
\begin{center}
\includegraphics[
height=2.4111in,
width=3.659in
]%
{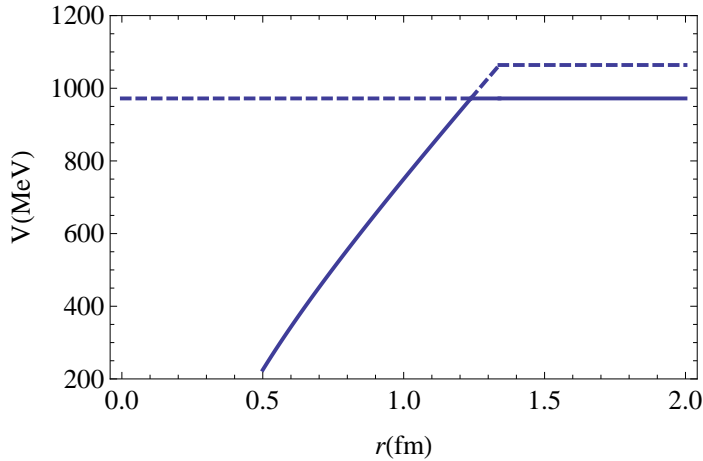}%
\caption{Generalized screened potential $V_{GSP}(r).$ The solid (dashed) line
indicates the potential in the first (second) energy region for $0^{+}%
(0^{++})$ $b\overline{b}$ states with $m_{b}=4793$ MeV, $\sigma=850$ MeV/fm,
$\chi=100$ MeV.fm, $M_{T_{1}}=10558$ MeV and $M_{T_{2}}=10650$ MeV (values of
the parameters and threshold masses from Section \ref{SIII}).}%
\label{Fig1a}%
\end{center}
\end{figure}

A lattice calculation of $E\left(  r\right)  $, implying the diagonalization
of a correlation matrix involving the bare valence $Q_{0}\overline{Q}_{0}$ and
meson $\left(  Q_{0}\overline{q}\right)  $ - meson $\left(  \overline{Q}%
_{0}q\right)  $ interacting configurations, has been carried out in reference
\cite{Bal05}. The results when only one meson $\left(  B\right)  $ - meson
$\left(  \overline{B}\right)  $ configuration, with mass $2m_{B},$ is
considered apart from the bare valence quark $\left(  Q_{0}\overline{Q}%
_{0}\right)  $\ are drawn in Fig. 22 of this reference that we reproduce here
for completeness as Fig.~\ref{Fig22prd05}%
\bigskip
The two thin curved lines following lattice data (circles and pentagons)
represent the calculated $E(r)-2m_{B}$ when only the meson-meson configuration
$B\overline{B}$ is implemented whereas the three thick lines correspond to an
educated guess for the case of $B\overline{B}$ and $B_{s}\overline{B_{s}}$
configurations. We should realize that in both cases $E(r)$ has, when not
close to any threshold, a Cornell type form.
\bigskip
It is important to emphasize that $E\left(  r\right)  $ from
Fig.~\ref{Fig22prd05} expresses the energy of two static color sources, $Q$
and $\overline{Q},$ implicitly incorporating screening effects, in terms of
the $Q-\overline{Q}$ distance. Notice that $Q$ can be interpreted as a dressed
valence quark (different from the bare valence quark $\left(  Q\neq
Q_{0}\right)  )$ since the $Q\overline{Q}$ interaction incorporates the effect
of meson $\left(  Q_{0}\overline{q}\right)  $ - meson $\left(  \overline
{Q}_{0}q\right)  $ configurations.
\bigskip
The Generalized Screened Potential Model (GSPM)\ is based on the assumption
that the dressed valence $Q\overline{Q}$ configuration represents, regarding
the spectrum, an effective description of a real meson. Then the meson masses
can be calculated from $Q\overline{Q}$ by solving the Schr\"{o}dinger equation
for a $Q-\overline{Q}$ Generalized Screened Potential (GSP) interaction.
\bigskip
In order to define the GSP let us start by defining from $E\left(  r\right)  $
an effective quark interaction potential as%
\[
V(r)\equiv E(r)-m_{Q}-m_{\overline{Q}}%
\]
where the masses $m_{Q}$ and $m_{\overline{Q}}$ are parameters to be fixed phenomenologically.
\bigskip
To analyze the form of $V(r)$ let us consider the more general two meson-meson
configuration case in Fig.~\ref{Fig22prd05}. Let us name the first (second)
threshold as $T_{1}$ $\left(  T_{2}\right)  $ (in Fig.~\ref{Fig22prd05}
$T_{1}=B\overline{B}$ $\left(  T_{2}=B_{s}\overline{B_{s}}\right)  .$ Let us
realize that the static approach implies that the two mesons forming the
threshold $T_{i}$ are in a relative $S-$ wave so that the threshold mass
$M_{T_{i}}$ corresponds to the sum of the masses of the mesons. Thus in
Fig.~\ref{Fig22prd05} $M_{T_{1}}=2m_{B}$ and $M_{T_{2}}=2m_{B_{s}}.$
\bigskip
As the forms of $E(r)$ are different below $M_{T_{1}},$ in between $M_{T_{1}}
$ and $M_{T_{2}},$ and above $M_{T_{2}},$ the potential $V(r)$ has different
forms in these energy regions. In this sense $V(r)$ is an energy dependent
potential. In practice this means that $Q\overline{Q}$ bound states with
masses $M_{Q\overline{Q}}$ belonging for example to the energy region
$0<M_{Q\overline{Q}}<$ $M_{T_{1}}$ should be obtained by solving the
Schr\"{o}dinger equation with the form of the potential corresponding to this
energy region and so on.
\bigskip
More precisely, in the first energy region defined by $0<M_{Q\overline{Q}%
}<M_{T_{1}}$, this is for $M_{Q\overline{Q}}\in\left[  M_{T_{0}},M_{T_{1}%
}\right]  $ where we have defined $M_{T_{0}}\equiv0$ in order to unify the
notation (let us realize that $T_{0}$ does not correspond to any real
meson-meson threshold), the form of the potential $V(r)$ will be called
$V_{\left[  M_{T_{0}},M_{T_{1}}\right]  }(r)$. This form is given by
$V_{\left[  M_{T_{0}},M_{T_{1}}\right]  }(r)=E(r)-m_{Q}-m_{\overline{Q}}$ with
$E(r)$ corresponding to the lower thick line in Fig.~\ref{Fig22prd05}.
According to the form of $E(r)$ when not close to threshold, this potential
has at short distances the Cornell type form $\left(  \sigma r-\frac{\chi}%
{r}+V_{0}\right)  .$ We shall include the constant $V_{0}$ in the definition
of the quark and antiquark masses so that we shall write the potential as
$\sigma r-\frac{\chi}{r}$. As can be checked from Fig.~\ref{Fig22prd05} this
form maintains up to a distance close below the crossing distance
$r_{T_{1}\text{ }}$defined from $V_{\left[  M_{T_{0}},M_{T_{1}}\right]
}(r_{T_{1}})=\sigma r_{T_{1}}-\frac{\chi}{r_{T_{1}}}=M_{T_{1}}-m_{Q}%
-m_{\overline{Q}}.$ Then $V_{\left[  M_{T_{0}},M_{T_{1}}\right]  }(r)$ starts
to flatten approaching its asymptotic value in this energy region $M_{T_{1}%
}-m_{Q}-m_{\overline{Q}}$ .
\bigskip
In the second energy region defined by $M_{T_{1}}<M_{Q\overline{Q}}<M_{T_{2}}
$ or $M_{Q\overline{Q}}\in\left[  M_{T_{1}},M_{T_{2}}\right]  $, the form of
the potential $V(r)$ will be called $V_{\left[  M_{T_{1}},M_{T_{2}}\right]
}(r).$ This form is given by $V_{\left[  M_{T_{1}},M_{T_{2}}\right]
}(r)=E(r)-m_{Q}-m_{\overline{Q}}$ with $E(r)$ corresponding to the
intermediate thick line in Fig.~\ref{Fig22prd05}. Therefore it is equal to
$M_{T_{1}}-m_{Q}-m_{\overline{Q}}$ from $r=0$ up to a distance close below
$r_{T_{1}\text{ }},$ then it rises until getting for a distance close above
$r_{T_{1}\text{ }}$ the form $\sigma r-\frac{\chi}{r}$. This form is
maintained up to a distance close below the crossing distance $r_{T_{2}\text{
}}$defined from $V_{\left[  M_{T_{1}},M_{T_{2}}\right]  }(r_{T_{2}})=\sigma
r_{T_{2}\text{ }}-\frac{\chi}{r_{T_{2}\text{ }}}=M_{T_{2}}-m_{Q}%
-m_{\overline{Q}}$ where $V_{\left[  M_{T_{1}},M_{T_{2}}\right]  }(r)$ starts
to flatten approaching its asymptotic value $M_{T_{2}}-m_{Q}-m_{\overline{Q}%
}.$
\bigskip
This analysis of the two threshold case can be easily generalized to the
general many threshold case by assuming that in between any two thresholds the
potential form is similar to $V_{\left[  M_{T_{1}},M_{T_{2}}\right]  }(r)$
through substitution of the corresponding thresholds.
\bigskip
For the sake of simplicity we shall reduce the size of the transition regions,
from the Cornell to the flat potentials, just to the crossing points
$r_{T_{i}}$. The Generalized Screened Potential (GSP) $V_{GSP}(r)$ is then
defined as:%
\begin{equation}
V_{GSP}(r)=V_{\left[  M_{T_{i-1}},M_{T_{i}}\right]  }%
(r)\text{\ \ \ \ \ if\ \ \ }M_{T_{i-1}}<M_{Q\overline{Q}}\leq M_{T_{i}}
\label{EDP}%
\end{equation}
with $i\geq1,$ and where the forms of the potential in the different spectral
regions are:%
\begin{equation}
V_{\left[  M_{T_{0}},M_{T_{1}}\right]  }(r)=\left\{
\begin{array}
[c]{c}%
\sigma r-\frac{\chi}{r}\text{ \ \ \ \ \ \ \ \ \ \ \ \ \ \ \ \ \ \ \ \ \ \ }%
r\leq r_{T_{1}}\\
\\
M_{T_{1}}-m_{Q}-m_{\overline{Q}}\text{ \ \ \ \ \ \ \ \ \ }r\geq r_{T_{1}}%
\end{array}
\right.  \label{pot1}%
\end{equation}
and%
\begin{equation}
V_{\left[  M_{T_{j-1}},M_{T_{j}}\right]  }(r)=\left\{
\begin{array}
[c]{c}%
M_{T_{j-1}}-m_{Q}-m_{\overline{Q}}\text{ \ \ \ \ \ }r\leq r_{T_{j-1}}\\
\\
\sigma r-\frac{\chi}{r}\text{\ \ \ \ \ \ \ \ \ \ \ }r_{T_{j-1}}\leq r\leq
r_{T_{j}}\\
\\
M_{T_{j}}-m_{Q}-m_{\overline{Q}}\text{ \ \ \ \ \ \ \ \ \ }r\geq r_{T_{j}}%
\end{array}
\right.  \label{pot12}%
\end{equation}
for $j>1$ with the crossing distances $r_{T_{j-1}}$ defined by
\begin{equation}
\sigma r_{T_{j-1}}-\frac{\chi}{r_{T_{j-1}}}=M_{T_{j-1}}-m_{Q}-m_{\overline{Q}}
\label{rti}%
\end{equation}
\bigskip
For instance the generalized screened potential $V_{GSP}(r)$ for
$b\overline{b}$ states with $I^{G}(J^{PC})=0^{+}(0^{++})$ quantum numbers,
whose first threshold is $B\overline{B}$, is drawn in Fig. \ref{Fig1a} for
the\ first and second energy regions.%
\bigskip
It is important to emphasize that $V_{GSP}(r)$ defined by (\ref{EDP}) is a
strictly confining potential (it always rises linearly from any threshold) so
that its spectrum only has $Q\overline{Q}$ bound states.
\section{Bottomonium \label{SIII}}
Bottomonium, made of heavy quarks $\left(  b\overline{b}\right)  $ is the
better framework, due to its non relativistic character, for the application
of the GSPM we have developed. One should keep in mind though that even in
this case the GSPM may be representing a rather simple approach to a real
meson description. On the one hand the model only incorporates screening from
meson - meson channels and no threshold widths have been taken into account.
Moreover the same effect from thresholds with $s\overline{s}$, $u\overline{u}$
or $d\overline{d}$ content has been considered but it could be different for
thresholds with $s\overline{s}$ content. On the other hand the Cornell
potential form of $V_{GSP}(r)$ when not close to any threshold, $\sigma
r-\frac{\chi}{r},$ does not contain spin dependent terms that, apart from
relativistic corrections, we know may give significant contributions to the
masses of the lower spectral states.
\bigskip
Anyhow, keeping in mind these possible shortcomings, we think it is worthwhile
to examine the physical consequences deriving from this simple dynamic model
for bottomonium to try to learn from them possible avenues for future progress.
\bigskip
In this Section we proceed to the calculation of the bottomonium spectrum. For
this purpose we fix first the values of the parameters of the model and we
list next the open flavor meson-meson threshold masses to be considered. Then
we detail the calculation of the spectrum for a particular case and compile
the bulk of results. From them the spectral effect of screening is analyzed.
\bigskip
\bigskip
\subsection{Parameters\label{SIIIA}}
To establish a criterion to fix the parameters $\sigma$, $\chi$ and $m_{Q}$
let us realize that in the first spectral region $\left[  M_{T_{0}},M_{T_{1}%
}\right]  $, for energies far below the first threshold, we hardly expect any
screening effect. In other words the Cornell potential
\begin{equation}
V_{Cor}(r)\equiv\sigma r-\frac{\chi}{r}\text{ \ \ \ }\left(  r:0\rightarrow
\infty\right)  \label{Cor}%
\end{equation}
should describe reasonably well this part of the spectrum. Actually this is
the case. It turns out that for a value of the Coulomb strength $\chi=100$
MeV.fm corresponding to a strong quark-gluon coupling $\alpha_{s}=\frac{3\chi
}{4\hbar}\simeq0.38$ (in agreement with the value derived from QCD from the
hyperfine splitting of $1p$ states in bottomonium \cite{Ynd95}), one can
choose correlated values of $\sigma$ and $m_{Q}$ to get such description. In
this regard, as we are dealing with a spin independent potential, we may
compare as usual the calculated $s-$ wave states with spin-triplets, the $p-$
wave states with the centroids obtained from data and the $d-$ wave states
with the only existing experimental candidates. Indeed it would be better a
comparison with the centroids for all states but the dearth of spin singlet
data makes this unfeasible.
\bigskip
Thus, by choosing for example $\sigma=850$ MeV/fm, a value within the
acceptable interval of values for the string tension in QCD, and $m_{b}=4793$
MeV, the differences from the calculated Cornell masses to data below the
first corresponding thresholds turn out to be less than $30$ MeV what
constitutes a reasonable overall description. We shall adopt these values so
that the set of parameters that will be used henceforth is%
\begin{equation}%
\begin{array}
[c]{c}%
\sigma=850\text{ MeV/fm}\\
\chi=100\text{ MeV.fm}\\
m_{b}=4793\text{ MeV}%
\end{array}
\label{valpar}%
\end{equation}
Let us advance that the degree of arbitrariness in the choice of the
parameters has no significant effect on the spectrum when they are required to
be correlated for a reasonable description of the lowest spectral states.
\subsection{$I(J^{PC})$ Thresholds\label{SIIIB}}
In order to apply the GSPM to a particular set of bottomonium states with
definite $I(J^{PC})$ we need the masses $M_{T_{i}}$ for meson ($Q_{0}%
\overline{q})$ - meson ($\overline{Q}_{0}q$) thresholds $\left(
q:u,d,s\right)  $ coupling to these quantum numbers. From these masses the
crossing radii $r_{T_{i}}$ are immediately calculated from (\ref{rti}).
\bigskip
Unfortunately not all thresholds are experimentally well known.
For example there is a known $0(1^{--})$ threshold from $B^{0}\overline
{B_{1}(5721)}^{0}$ where $B_{1}(5721)^{0}$ is a $\left(  ^{3}P_{1}-^{1}%
P_{1}\right)  $ mixing state (see for example \cite{PDG12}). However the
$\left(  ^{3}P_{1}-^{1}P_{1}\right)  $ partner of $B_{1}(5721)^{0}$ is not
known yet although we expect this missing state to have a mass close to that
of $B_{1}(5721)^{0}$. We shall call it $B_{1}(?).$ Therefore the mass of the
threshold $B^{0}\overline{B_{1}\left(  ?\right)  }^{0}$ is not well known.
In other cases the situation is reversed since a threshold mass is known but
its quantum numbers are not well established. This is for example the
situation for thresholds including the meson $B_{J}^{\ast}(5732)$ that we
shall tentatively assign to $J^{P}=0^{+}.$
\bigskip
The list of thresholds for bottomonium, with their corresponding masses and
crossing radii, appear in Tables~\ref{tab2++} and \ref{tab2--}. The lack of
knowledge about further thresholds prevents extending the list to higher energies.
\begin{table}[ptb]%
\begin{tabular}
[c]{ccccc}%
$I(J^{PC})$ & $T_{i}$ & $%
\begin{array}
[c]{c}%
\text{Bottomonium}\\
\text{Thresholds}%
\end{array}
$ & $%
\begin{array}
[c]{c}%
M_{T_{i}}\\
\text{(MeV)}%
\end{array}
$ & $%
\begin{array}
[c]{c}%
r_{T_{i}}\\
\text{(fm)}%
\end{array}
$\\\hline
&  &  &  & \\
$0(0^{++})$ &  &  &  & \\
& $T_{1}$ & $\left(  B^{0}\overline{B}^{0},B^{+}B^{-}\right)  _{I=0}$ &
10558 & 1.24\\
& $T_{2}$ & $\left(  B^{\ast0}\overline{B^{\ast}}^{0},B^{\ast+}B^{\ast
-}\right)  _{I=0}$ & 10650 & 1.34\\
& $T_{3}$ & $B_{s}^{0}\overline{B_{s}}^{0}$ & 10734 & 1.43\\
&  &  &  & \\
$0(1^{++})$ &  &  &  & \\
& $T_{1}$ & $(B^{0}\overline{B^{\ast}}^{0},B^{+}\overline{B^{\ast}}^{-}%
)_{I=0}+c.c.$ & 10604 & 1.29\\
& $T_{2}$ & $B_{s}^{0}\overline{B_{s}^{\ast}}+c.c.$ & $10782$ & 1.49\\
&  &  &  & \\
$0(2^{++})$ &  &  &  & \\
& $T_{1}$ & $\left(  B^{\ast0}\overline{B^{\ast}}^{0},B^{\ast+}B^{\ast
-}\right)  _{I=0}$ & 10650 & 1.34\\
& $T_{2}$ & $B_{s}^{\ast}\overline{B_{s}^{\ast}}$ & 10830 & 1.54\\
&  &  &  &
\end{tabular}
\caption{Open flavor meson-meson thresholds for $0\left(  J^{++}\right)  $
$b\overline{b}$ states. Threshold masses $\left(  M_{T_{i}}\right)  $ obtained
from the bottom and bottom strange meson masses quoted in \cite{PDG12}.
Crossing distances $\left(  r_{T_{i}}\right)  $ calculated from (\ref{rti}).}%
\label{tab2++}%
\end{table}\begin{table}[ptbptb]%
\begin{tabular}
[c]{ccccc}%
$I(J^{PC})$ & $T_{i}$ & $%
\begin{array}
[c]{c}%
\text{Bottomonium}\\
\text{Thresholds}%
\end{array}
$ & $%
\begin{array}
[c]{c}%
M_{T_{i}}\\
\text{(MeV)}%
\end{array}
$ & $%
\begin{array}
[c]{c}%
r_{T_{i}}\\
\text{(fm)}%
\end{array}
$\\\hline
&  &  &  & \\
$0(1^{--})$ &  &  &  & \\
& $T_{1}$ & $%
\begin{array}
[c]{c}%
(B^{0}\overline{B_{1}(5721)}^{0},\\
B^{+}B_{1}(5721)^{-})_{I=0}-c.c.\\
\\
(B^{0}\overline{B_{1}(?)}^{0},\\
B^{+}B_{1}(?)^{-})_{I=0}-c.c.
\end{array}
$ & $%
\begin{array}
[c]{c}%
11003\\
\\
?
\end{array}
$ & $%
\begin{array}
[c]{c}%
1.73\\
\\
?
\end{array}
$\\
&  &  &  & \\
&  &  &  & \\
&  &  &  & \\
& $T_{2}$ & $%
\begin{array}
[c]{c}%
(B^{\ast0}\overline{B_{0}^{\ast}(5732)}^{0},\\
B^{\ast+}B_{0}^{\ast}(5732)^{-})_{I=0}-c.c.
\end{array}
$ & $11023$ & $1.76$\\
&  &  &  & \\
&  &  &  & \\
&  &  &  & \\
& $T_{3}$ & $%
\begin{array}
[c]{c}%
(B^{\ast0}\overline{B_{1}(5721)}^{0},\\
B^{\ast+}B_{1}(5271)^{-})_{I=0}-c.c.\\
\\
(B^{\ast0}\overline{B_{1}(?)}^{0},\\
B^{\ast+}B_{1}(?)^{-})_{I=0}-c.c.
\end{array}
$ & $%
\begin{array}
[c]{c}%
11049\\
\\
?
\end{array}
$ & $%
\begin{array}
[c]{c}%
1.79\\
\\
?
\end{array}
$\\
&  &  &  & \\
&  &  &  & \\
&  &  &  & \\
& $T_{4}$ & $%
\begin{array}
[c]{c}%
(B^{\ast0}\overline{B_{2}^{\ast}(5747)}^{0},\\
B^{\ast+}B_{2}^{\ast}(5747)^{-})_{I=0}-c.c.
\end{array}
$ & $11072$ & $1.81$%
\end{tabular}
\caption{Open flavor meson-meson thresholds for $0(1^{--})$ $b\overline{b}$
states. Threshold masses $\left(  M_{T_{i}}\right)  $ calculated from the
bottom and bottom strange meson masses quoted in \cite{PDG12}. Crossing
distances $\left(  r_{T_{i}}\right)  $ calculated from (\ref{rti}). For
$B_{J}^{\ast}(5732)$ with quoted mass $5691$ MeV we have assumed $J=0.$ A
question mark has been used for the mass of an unknown meson and the mass of
the corresponding threshold. }%
\label{tab2--}%
\end{table}
\bigskip
It is important to remark that we have used isospin symmetry to construct
thresholds with well defined isospin. This means that we are neglecting the
mass differences between the electrically neutral and charged members of the
same isospin multiplet, for example $B^{0}$ and $B^{\pm}$ with PDG quoted
masses \cite{PDG12} $5279.53\pm0.33$ MeV and $5279.15\pm0.31$ MeV respectively.
\bigskip
Regarding the $C$ parity for a threshold formed by two mesons $\mathfrak{M}%
_{1}$ and $\mathfrak{M}_{2}$ we can construct the combinations $\left(
\mathfrak{M}_{1}\mathfrak{M}_{2}\pm c.c\right)  $ with $C$ parity $+$ and $-$
respectively. Notice though that if $\mathfrak{M}_{2}=\overline{\mathfrak{M}%
_{1}}$ then, as the two mesons are in a relative $S-$ wave, we have
$\overline{\mathfrak{M}_{1}}\mathfrak{M}_{1}=\left(  -\right)  ^{j_{1}%
+j_{1}-j}\mathfrak{M}_{1}\overline{\mathfrak{M}_{1}}$ where $j_{1}$ stands for
the spin of $\mathfrak{M}_{1}$ and $j$ for the total spin of the threshold.
Therefore only one combination in $\mathfrak{M}_{1}\overline{\mathfrak{M}_{1}%
}\pm c.c$ is allowed for a given value of $j$ (the other vanishes)$.$ For
example the $I=0$ threshold $B^{\ast}\overline{B^{\ast}}$ with $j_{1}=1$ has
positive $C$ parity when coupled to $j=0,2.$
\subsection{Spectrum\label{SIIIC}}
Bottomonium states are obtained by solving the Schr\"{o}dinger equation for
the GSP potential $V_{GSP}(r).$ In the energy region $\left[  M_{T_{i-1}%
},M_{T_{i}}\right]  $ they satisfy%
\begin{align}
&  \left(  \mathcal{T}+V_{\left[  M_{T_{i-1}},M_{T_{i}}\right]  }\right)
\left\vert (Q\overline{Q})_{k_{\left[  T_{i-1},T_{i}\right]  }}\right\rangle
\label{SchEQM}\\
&  =M_{k_{\left[  T_{i-1},T_{i}\right]  }}\left\vert (Q\overline
{Q})_{k_{\left[  T_{i-1},T_{i}\right]  }}\right\rangle \nonumber
\end{align}
where $\mathcal{T}$\ stands for the kinetic energy operator, $\left\vert
(Q\overline{Q})_{k_{\left[  T_{i-1},T_{i}\right]  }}\right\rangle $ for the
bound state and $M_{k_{\left[  T_{i-1},T_{i}\right]  }}$ for its mass. As we
have a radial potential we use the spectroscopic notation $k\equiv nl,$ in
terms of the radial, $n,$ and orbital angular momentum, $l,$ quantum numbers
of the $Q\overline{Q}$ system.
\bigskip
To fix the ideas let us consider for example the spectral states for
$0^{+}(0^{++})$ $b\overline{b}.$
In the first energy region the potential $V_{\left[  M_{T_{0}},M_{T_{1}%
}\right]  }(r),$ given by (\ref{pot1}), reads (solid line in Fig.
\ref{Fig1a})
\[
V_{\left[  0,10558\right]  }(r)=\left\{
\begin{array}
[c]{c}%
\sigma r-\frac{\chi}{r}\text{ \ \ \ \ \ \ \ \ \ \ \ \ \ \ \ \ }r\leq1.24\text{
fm}\\
\\
972\text{ MeV\ \ \ \ \ \ \ \ \ \ \ \ \ \ }r\geq1.24\text{ fm}%
\end{array}
\right.
\]
where $M_{T_{1}}$ and $r_{T_{1}}$ have been taken from Table \ref{tab2++} and
the values of the parameters $\left(  \sigma,\chi,m_{b}\right)  $ are given by
(\ref{valpar}).
\bigskip
By solving the Schr\"{o}dinger equation for $V_{\left[  0,10558\right]  }(r)$
we get the GSPM spectrum in $\left[  M_{T_{0}},M_{T_{1}}\right]  $. It has
only three bound states states, $1p_{\left[  T_{0},T_{1}\right]  },$
$2p_{\left[  T_{0},T_{1}\right]  }$ and $3p_{\left[  T_{0},T_{1}\right]  },$
whose masses $M_{k_{\left[  T_{0},T_{1}\right]  }}$ generically denoted by
$M_{GSP}$ are listed in Table~\ref{tab0++}.
\bigskip
In the second energy region the potential, $V_{\left[  M_{T_{1}},M_{T_{2}%
}\right]  }(r),$ reads (dashed line in Fig. \ref{Fig1a})
\[
V_{\left[  10558,10650\right]  }(r)=\left\{
\begin{array}
[c]{c}%
972\text{ MeV\ \ \ \ \ \ \ \ }r\leq1.24\text{ fm}\\
\\
\sigma r-\frac{\chi}{r}\text{\ \ \ \ \ }1.24\text{ fm}\leq r\leq1.34\text{
fm}\\
\\
1064\text{ MeV\ \ \ \ \ \ \ \ }r\geq1.34\text{ fm}%
\end{array}
\right.
\]
where the threshold masses and crossing radii are taken from Table
\ref{tab2++}. The spectrum has only one bound state $1p_{\left[  T_{1}%
,T_{2}\right]  }$ whose mass $M_{1p_{\left[  T_{1},T_{2}\right]  }}$
generically denoted by $M_{GSP}$ is listed in Table~\ref{tab0++}.
\begin{table}[ptb]%
\begin{tabular}
[c]{ccccc}%
$%
\begin{array}
[c]{c}%
b\overline{b}\\
0^{+}(0^{++})
\end{array}
$ & $\left[  T_{i-1},T_{i}\right]  $ & $%
\begin{array}
[c]{c}%
\left[  M_{T_{i-1}},M_{T_{i}}\right] \\
\text{MeV}%
\end{array}
$ & $%
\begin{array}
[c]{c}%
\text{GSPM}\\
\text{States}\\
k_{\left[  T_{i-1},T_{i}\right]  }%
\end{array}
$ & $%
\begin{array}
[c]{c}%
M_{GSP}\\
\text{MeV}%
\end{array}
$\\\hline
&  &  &  & \\
& $\left[  T_{0},T_{1}\right]  $ & $\left[  0,10558\right]  $ & $1p_{\left[
T_{0},T_{1}\right]  }$ & $9920$\\
&  &  & $2p_{\left[  T_{0},T_{1}\right]  }$ & $10259$\\
&  &  & $3p_{\left[  T_{0},T_{1}\right]  }$ & $10521$\\
&  &  &  & \\
&  &  &  & \\
& $\left[  T_{1},T_{2}\right]  $ & $\left[  10558,10650\right]  $ &
$1p_{\left[  T_{1},T_{2}\right]  }$ & $10620$\\
&  &  &  &
\end{tabular}
\caption{Calculated $0^{+}(0^{++})$ $b\overline{b}$ masses from $V_{GSP}(r),$
generically denoted by $M_{GSP},$ in the first two energy regions indicated by
the thresholds $\left[  T_{i-1},T_{i}\right]  $ and their masses $\left[
M_{T_{i-1}},M_{T_{i}}\right]  $. }%
\label{tab0++}%
\end{table}
\bigskip
By proceeding in the same way for higher energy regions and for different
quantum numbers we get the complete GSP bound state spectrum.
\bigskip
The spectrum for $0^{+}(J^{++})$ $b\overline{b}$ states from the generalized
screened potential $V_{GSP}(r)$ given by (\ref{EDP}) is shown in
Table~\ref{tab13}. The spectrum from the Cornell potential $V_{Cor}(r)$ given
by (\ref{Cor}) with the same values of the parameters $\sigma$, $\chi$ and
$m_{Q}$ given by (\ref{valpar}) is also listed for comparison.
\begin{table}[ptb]%
\begin{tabular}
[c]{ccccc}%
$J^{PC}$ & $%
\begin{array}
[c]{c}%
\text{GSP}\\
\text{States}\\
k_{\left[  T_{i-1},T_{i}\right]  }%
\end{array}
$ & $%
\begin{array}
[c]{c}%
M_{EQM}\\
\text{MeV}%
\end{array}
$ & $%
\begin{array}
[c]{c}%
M_{PDG}\\
\text{MeV}%
\end{array}
$ & $%
\begin{array}
[c]{c}%
M_{Cor}\left(  k\right) \\
\text{MeV}%
\end{array}
$\\\hline
&  &  &  & \\
$0^{++}$ & $1p_{\left[  T_{0},T_{1}\right]  }$ & $9920$ & $9859.44\pm
0.42\pm0.31$ & $9920$ $\left(  1p\right)  $\\
$1^{++}$ & $1p_{\left[  T_{0},T_{1}\right]  }$ & $9920$ & $9892.78\pm
0.26\pm0.31$ & $9920$ $\left(  1p\right)  $\\
$2^{++}$ & $1p_{\left[  T_{0},T_{1}\right]  }$ & $9920$ & $9912.21\pm
0.26\pm0.31$ & $9920$ $\left(  1p\right)  $\\
$0^{++}$ & $2p_{\left[  T_{0},T_{1}\right]  }$ & $10259$ & $10232.5\pm
0.4\pm0.5$ & $10259$ $\left(  2p\right)  $\\
$1^{++}$ & $2p_{\left[  T_{0},T_{1}\right]  }$ & $10259$ & $10255.46\pm
0.22\pm0.50$ & $10259$ $\left(  2p\right)  $\\
$2^{++}$ & $2p_{\left[  T_{0},T_{1}\right]  }$ & $10259$ & $10268.65\pm
0.22\pm0.50$ & $10259$ $\left(  2p\right)  $\\
$0^{++}$ & $3p_{\left[  T_{0},T_{1}\right]  }$ & $10521$ &  & $10531$ $\left(
3p\right)  $\\
$1^{++}$ & $3p_{\left[  T_{0},T_{1}\right]  }$ & $10526$ &  & $10531$ $\left(
3p\right)  $\\
&  &  & $10530\pm5\pm9$ & \\
$2^{++}$ & $3p_{\left[  T_{0},T_{1}\right]  }$ & $10528$ &  & $10531$ $\left(
3p\right)  $\\
&  &  &  & \\
$0^{++}$ & $1p_{\left[  T_{1},T_{2}\right]  }$ & $10620$ &  & \\
$1^{++}$ & $1p_{\left[  T_{1},T_{2}\right]  }$ & $10668$ &  & \\
$0^{++}$ & $1p_{\left[  T_{2},T_{3}\right]  }$ & $10704$ &  & \\
$2^{++}$ & $1p_{\left[  T_{1},T_{2}\right]  }$ & $10710$ &  & \\
&  &  &  & $10768$ $\left(  4p\right)  $\\
$1^{++}$ & $2p_{\left[  T_{1},T_{2}\right]  }$ & $10776$ &  & \\
$0^{++}$ & $1p_{\left[  T_{3},T_{4}\right]  }$ & $10784$ &  & \\
$2^{++}$ & $2p_{\left[  T_{1},T_{2}\right]  }$ & $10815$ &  & \\
&  &  &  &
\end{tabular}
\caption{Calculated $J^{++}$ bottomonium masses from $V_{GSP}(r):M_{GSP}.$
Masses for experimental resonances, $M_{PDG},$ have been taken from
\cite{PDG12}. For $p$ waves we quote separately the $np_{0}$, $np_{1} $ and
$np_{2}$ states. Masses and states from the Cornell potential $V_{Cor}(r),$
denoted by $M_{Cor}$ $\left(  k\right)  $ are also shown for comparison.}%
\label{tab13}%
\end{table}
\bigskip
For $0^{-}\left(  1^{--}\right)  $ $b\overline{b}$ states there is some
uncertainty in the calculation of the spectrum from the unknown threshold
masses. Moreover, the possible accumulative effect of almost degenerate
thresholds is out of the scope of the GSP such as has been defined. From this
uncertainty we can not reasonably determine the spectrum around and above
$11000$ MeV. Hence we limit our calculation to the first energy region having
taken the first threshold mass at $11003$ MeV. In Table~\ref{tab12} we list
these results as well as the ones from the Cornell potential, with the same
values of the parameters $\sigma$, $\chi$ and $m_{Q},$ for comparison.
\begin{table}[ptb]%
\begin{tabular}
[c]{ccccc}%
$J^{PC}$ & $%
\begin{array}
[c]{c}%
\text{EQM}\\
\text{States}\\
k_{\left[  T_{i-1},T_{i}\right]  }%
\end{array}
$ & $%
\begin{array}
[c]{c}%
M_{EQM}\\
\text{MeV}%
\end{array}
$ & $%
\begin{array}
[c]{c}%
M_{PDG}\\
\text{MeV}%
\end{array}
$ & $%
\begin{array}
[c]{c}%
M_{Cor}\left(  k\right) \\
\text{MeV}%
\end{array}
$\\\hline
&  &  &  & \\
$1^{--}$ & $1s_{\left[  T_{0},T_{1}\right]  }$ & $9459$ & $9460.30\pm0.26$ &
$9459$ $\left(  1s\right)  $\\
& $2s_{\left[  T_{0},T_{1}\right]  }$ & $10012$ & $10023.026\pm0.31$ & $10012$
$\left(  2s\right)  $\\
& $1d_{\left[  T_{0},T_{1}\right]  }$ & $10157$ & $10163.7\pm1.4$ & $10157$
$\left(  1d\right)  $\\
& $3s_{\left[  T_{0},T_{1}\right]  }$ & $10342$ & $10355.2\pm0.5$ & $10342$
$\left(  3s\right)  $\\
& $2d_{\left[  T_{0},T_{1}\right]  }$ & $10438$ &  & $10438$ $\left(
2d\right)  $\\
& $4s_{\left[  T_{0},T_{1}\right]  }$ & $10608$ & $10579.4\pm1.2$ & $10608$
$\left(  4s\right)  $\\
& $3d_{\left[  T_{0},T_{1}\right]  }$ & $10682$ &  & $10682$ $\left(
3d\right)  $\\
& $5s_{\left[  T_{0},T_{1}\right]  }$ & $10840$ &  & $10841$ $\left(
5s\right)  $\\
&  &  & $10876\pm11$ & \\
& $4d_{\left[  T_{0},T_{1}\right]  }$ & $10899$ &  & $10902$ $\left(
4d\right)  $\\
&  &  &  &
\end{tabular}
\caption{Calculated $1^{--}$ bottomonium masses from $V_{GSP}(r):M_{GSP}.$
Masses for experimental resonances, $M_{PDG},$ have been taken from
\cite{PDG12}. Masses and states from the Cornell potential $V_{Cor}(r),$
denoted by $M_{Cor}\left(  k\right)  $ are also shown for comparison.}%
\label{tab12}%
\end{table}
\subsection{Screening Effects\label{SIIID}}
A look at Table~\ref{tab13} makes clear that the more significant spectral
effect from the generalized screened potential $V_{GSP}(r)$ is the bigger
number of spectral states above the first meson-meson threshold as compared to
the non screened Cornell potential $V_{Cor}(r)$ case. Thus, for example there
are three $0^{+}(0^{++})$ $b\overline{b}$ GSPM bound states with masses
$\left(  10620,10704,10784\right)  $ MeV between $10558$ MeV, the mass of the
first threshold, and $10830$ MeV, the mass of the last known threshold, for
only one Cornell state with mass $10768$ MeV in this energy interval.
\bigskip
Regarding the spectrum in the first energy region is almost identical for both
potentials, the only difference being a slightly bigger attraction for
$V_{GSP}(r)$ which makes the states close below threshold to be lower in mass
than the corresponding Cornell ones. Notice that this extra attraction could
make in some particular case that a state that is close above threshold for
the Cornell potential lies close below threshold for the screened potential.
\bigskip
An additional effect from the screened potential is the breaking of the
$J^{++}=\left(  0,1,2\right)  ^{++}$ degeneracy implied by the Cornell
potential. This is due to the different values of the threshold masses in each
case. However, at the level of precision of our calculation, we obtain for the
masses of the $1p\left(  0,1,2\right)  ^{++}$ the same value ($9919.6$ MeV).
This has to do with the fact that these states lying quite below the first
threshold are very little affected by it. A similar argument applies to the
$2p\left(  0,1,2\right)  ^{++}$ states with calculated masses ($10258.5$ MeV,
$10258.6$ MeV, $10258.6$ MeV), rounded off to $10259$ MeV in Table~\ref{tab13}%
. One should not forget though that a a more important contribution to this
breaking may come from the non considered spin-dependent terms in the potential.
\bigskip
\bigskip
Therefore we may conclude that a denser spectral pattern than conventionally
considered is the main feature resulting from the application of the GSPM. In
other words screening effects in the way we have implemented them give rise to
the appearance of new spectral states (not present in the non screened
potential case). This can be understood if we think of an alternative (but
much more complicated technically) equivalent method for calculating the
spectrum based on the consideration of interacting bare valence $Q_{0}%
\overline{Q}_{0}$ and meson $\left(  Q_{0}\overline{q}\right)  $ - meson
$\left(  \overline{Q}_{0}q\right)  $ configurations. Then it is clear that
through configuration mixing more spectral states than the pure Cornell (non
screened) states corresponding to $Q_{0}\overline{Q}_{0}$ are present.
\bigskip
Unfortunately we have not yet enough data to validate or refute this
conclusion. A scan for $J^{++}$ states in the mentioned energetic region (from
$10558$ MeV to $10830$ MeV) could shed definite light about this prediction.
With respect to this it should be mentioned that most of the new spectral
states are related to at least one threshold with $s\overline{s}$ content what
could imply a reduction of the formation probability for them. The only
exception is the $0^{+}(0^{++})(10620)$ which is a priory the ideal candidate
to check the GSPM.
\
\section{Summary\label{SIV}}
A new nonrelativistic quark model to study the spectrum of heavy quark mesons
has been developed. The model is built in terms of effective quark degrees of
freedom interacting through a potential that incorporates screening effects
from meson-meson configurations. The form of this interaction potential that
we call Generalized Screened Potential, or abbreviate GSP, has been proposed
from lattice results and exhibits a characteristic dependence on the energy
interval of application.
\bigskip
The model, called Generalized Screened Potential Model, or abbreviate GSPM,
has been applied to calculate the bottomonium spectrum (the only non
relativistic meson system). A richer spectrum (bigger number of bound states)
than the one resulting from the non-screened Cornell potential is obtained. In
particular extra $J^{++}$ bottomonium states above the first meson-meson
threshold appear. Certainly the masses of these new states may be shifted when
dynamic corrections are implemented. However as the form of the potential in
between two thresholds is determined to a large extent by the threshold masses
(indeed it can be approximated by a spherical well) we hardly expect any
change in the number of calculated states when these corrections are
incorporated. Therefore we consider the presence of these extra states a quite
robust distinctive prediction of the model.
\bigskip
It should be emphasized that by construction the model is suited for the
calculation of spectral masses. In this regard it represents a very simple and
efficient alternative to a couple channel calculation of the spectrum
involving bare valence and meson - meson configurations. As a counterpart a
quantitative treatment of decay processes (for example strong decays to open
flavor mesons) may require the explicit consideration of meson - meson
configurations as in a couple channel scheme.
\bigskip
The generalization of the Generalized Screened Potential Model to other meson
sectors is feasible but some dynamic implementations may be required. In
particular the analysis of charmonium with a richer spectrum than
conventionally expected deserves special attention and will be the subject of
future work.\
\bigskip
In summary we have proposed a spectral Generalized Screened Potential Model
which can be considered as a first attempt to incorporate lattice screening
effects from meson-meson thresholds within a non relativistic quark model framework.
\bigskip
\bigskip
This work has been supported by HadronPhysics2, by Ministerio de Econom\'{\i}a
y Competitividad (Spain) and UE FEDER grant FPA2010-21750-C02-01, and by GV Prometeo2009/129.

\end{document}